\documentclass[10pt,twoside]{hsqcd}
\usepackage{epsf,amsmath}
\usepackage{graphicx}

\begin{document}

\title{SEARCHES FOR NEW PHYSICS AT HERA}
\author{D.~M.~SOUTH \\
{\it on behalf of the H1 and ZEUS collaborations} \\
Deutsches Elektronen Synchrotron\\
Notkestrasse 85, 22607, Hamburg, Germany\\
E-mail: David.South@desy.de }

\maketitle

\vspace{0.5cm}

\begin{abstract}
\noindent The latest results from the H1 and ZEUS collaborations
are presented on rare Standard Model processes and searches for
physics Beyond the Standard Model. Intriguing events containing high
transverse momentum leptons are observed by H1 and many competitive
limits are set by both collaborations on the production of new particles
in a variety of supersymmetry scenarios.
 
\end{abstract}

\markboth{\large \sl D.~M.~SOUTH
\hspace*{2cm} HSQCD 2005} {\large \sl \hspace*{1cm} SEARCHES FOR NEW PHYSICS AT HERA}

\section{Introduction}

A comprehensive physics programme is employed by the H1 and ZEUS
experiments at the HERA $ep$ collider. Together with measuring the structure
of the proton, the deep inelastic collisions (DIS) produced at HERA, at a centre
of mass up to 318~GeV, provide an ideal environment to study rare processes,
set constraints on the Standard Model (SM) and search for new particles
and physics beyond the Standard Model (BSM).

\section{Rare Standard Model Processes}

\subsection{Events containing Isolated Leptons and Missing Transverse Momentum}

Events containing a high transverse momentum ($P_{T}$) isolated electron or muon and large
missing transverse momentum have been observed at HERA
\cite{isoleph1origwpaper,isolepzeusorigwpaper,isoleph1newwpaper}. The main SM contribution
to such a topology comes from the production of real $W$ bosons $ep \rightarrow eW^{\pm}X$
with subsequent leptonic decay $W \rightarrow l\nu$. An excess of HERA~I (1994--2000)
data events compared to the SM prediction was reported by the H1 collaboration in
\cite{isoleph1newwpaper}, which was not confirmed by the ZEUS collaboration, although
using a slightly different analysis approach \cite{zeustop}.

\begin{figure}
\begin{minipage}{.51\linewidth}
  \centering
  \includegraphics[width=.95\textwidth]{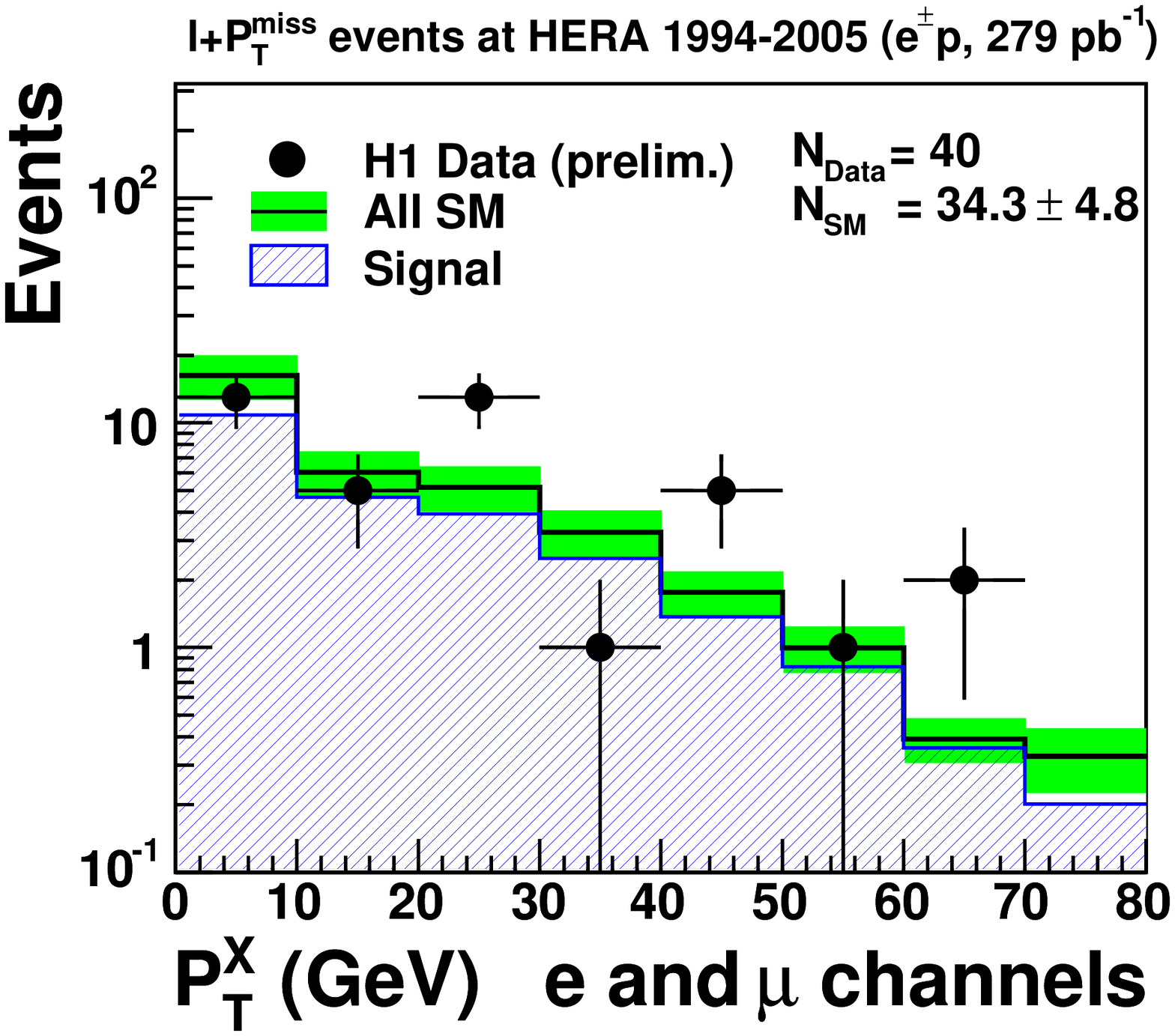}
  \caption{The hadronic transverse momentum spectrum of the observed events in the complete
	H1 HERA $e^{\pm}p$ data sample (${\cal L} =$~279~pb$^{-1}$) compared to the
	SM expectation \cite{isoleph1hera2}.}
\label{fig:isolep}
\end{minipage}
\hspace{0.1cm}
\begin{minipage}{.45\linewidth}
  \centering
  \includegraphics[width=.95\textwidth]{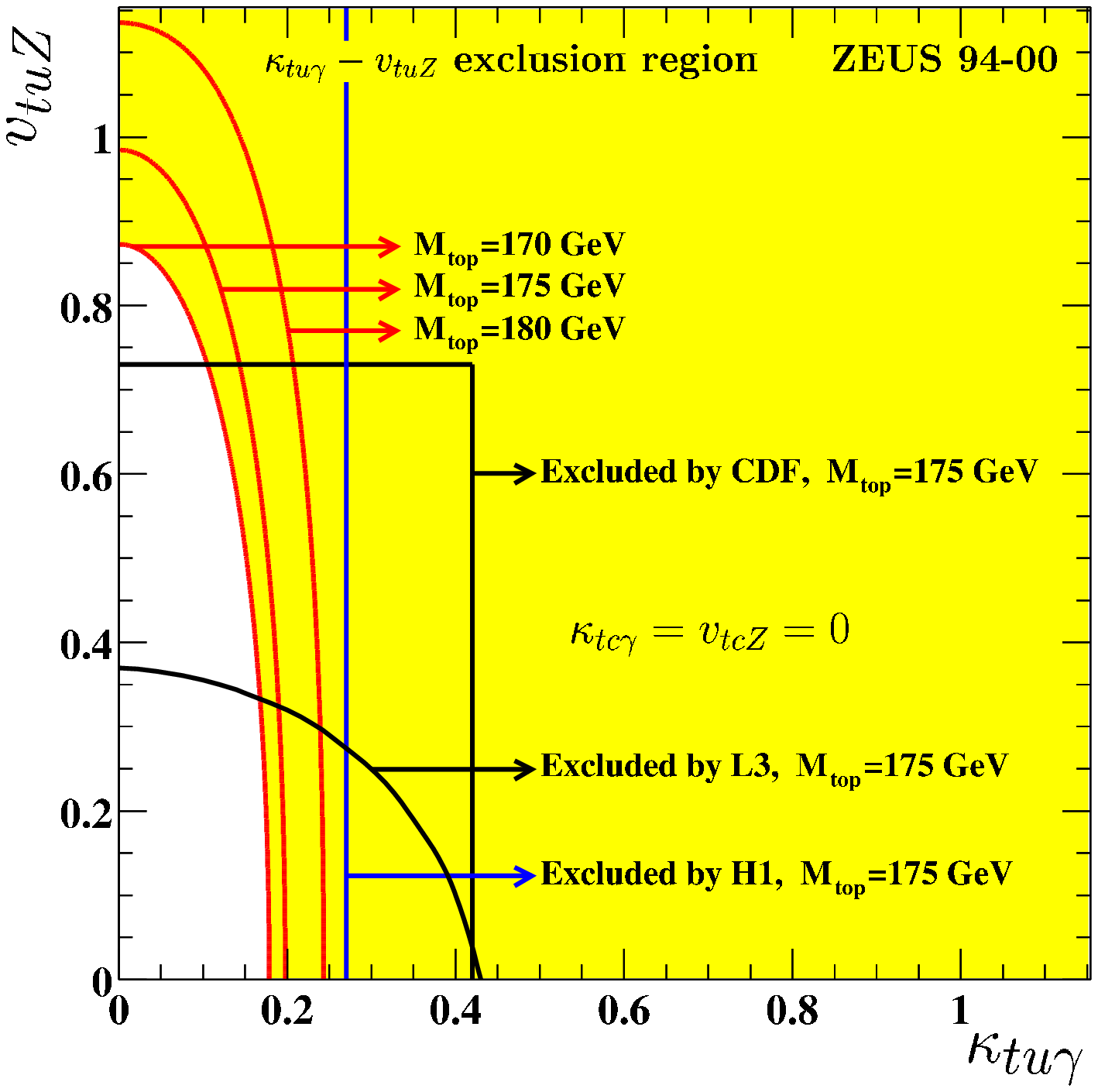}
  \caption{The ZEUS \cite{zeustop} and H1 \cite{h1top} exclusion limits at 95\% CL
	on the anomalous $\kappa_{tu\gamma}$ coupling compared to limits derived
	by the CDF \cite{cdftop} and the L3 \cite{leptop} experiments.}
\label{fig:singletop}
\end{minipage}
\end{figure}

The H1 analysis has been updated to include new $e^{\pm}p$ data from the ongoing HERA~II phase
(since 2003), resulting in a total analysed luminosity of 279~pb$^{-1}$ \cite{isoleph1hera2}.
A total of 40 events are observed in the data, compared to a SM prediction of
34.3~$\pm$~4.8\footnote{These numbers are updated with respect to those presented at
St.~Petersburg; for full details see the second reference in \cite{isoleph1hera2}.}. At
large values of hadronic transverse momentum $P_{T}^{X}$ an excess of data events is observed
over the SM expectation, as can be seen in figure \ref{fig:isolep}. Additionally, the observed
H1 excess is only present in the $e^{+}p$ data, where for $P_{T}^{X} >$~25~GeV a total of 15
data events are observed compared to a SM prediction of 4.6~$\pm$~0.8. A re-analysis of the
ZEUS electron channel has been performed \cite{isolepzeushera2}, using $e^{+}p$ data from
1999-2000 in addition to HERA~II $e^{+}p$ data from 2003-04 (total luminosity 106~pb$^{-1}$),
leading to similar backgrounds as in the H1 analysis. However, the ZEUS analysis does not
confirm the H1 excess, where only one electron candidate is observed in the
region $P_{T}^{X} >$~25~GeV.

Both experiments have also searched for the non--standard production of single top quarks at
HERA~I via the Flavour Changing Neutral Current (FCNC) process, where the analysis strategy
is an extension of the isolated lepton analysis \cite{zeustop,h1top}. Some of the H1 events at
large $P_{T}^{X}$ show kinematics compatible with the single top hypothosis, but no signal
can be claimed. Figure \ref{fig:singletop} shows limits derived at 95\% confidence level
(CL) by both experiments on the anomalous FCNC $\kappa_{tu\gamma}$ coupling. It can be seen
that HERA has the best sensitivity to the $\kappa_{tu\gamma}$ coupling in the region
where $v_{tuZ}$ is small.

\subsection{Multi-lepton Events}

\begin{figure}[h]
  \centering
  \includegraphics[width=.44\textwidth]{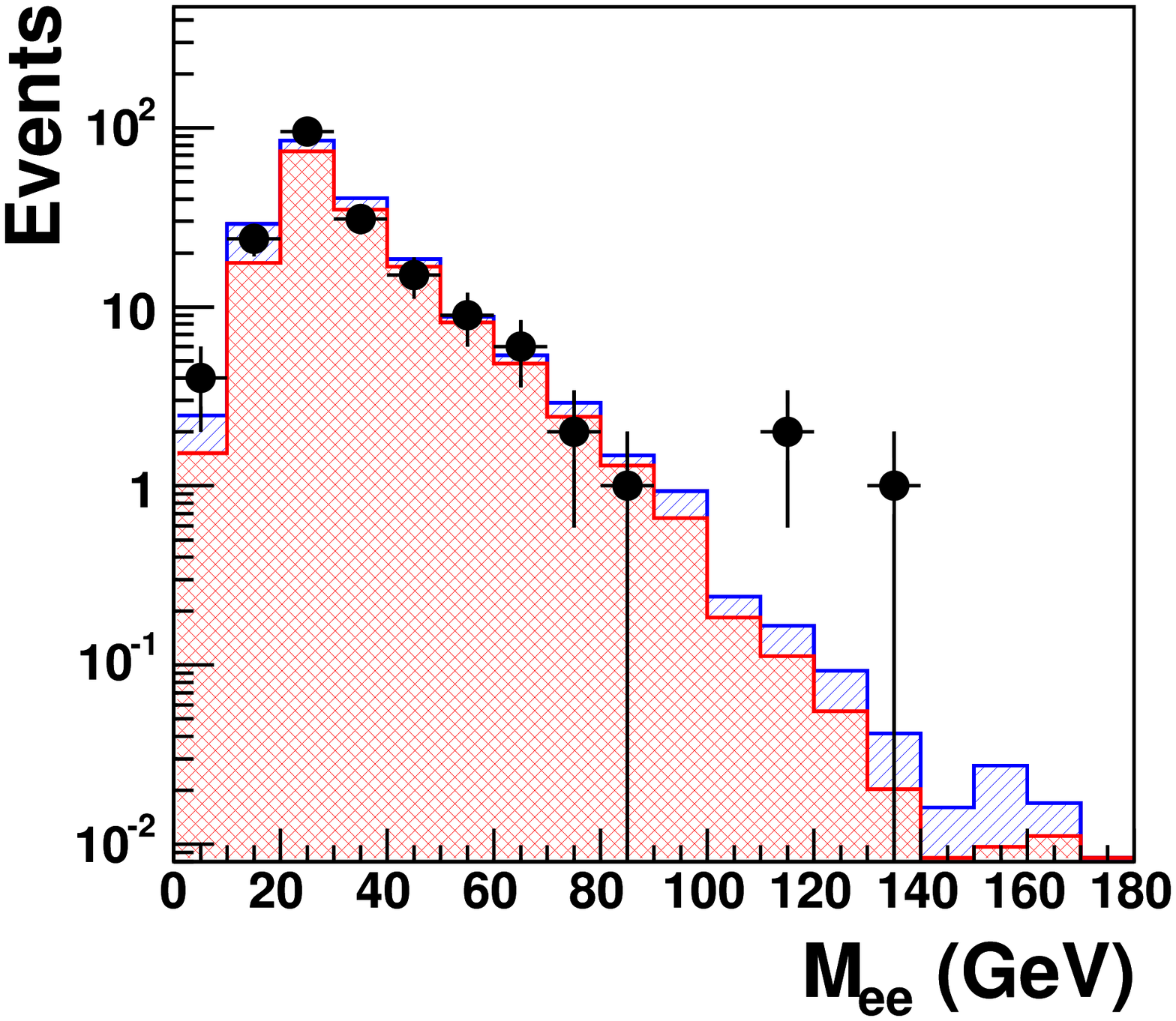}
  \includegraphics[width=.44\textwidth]{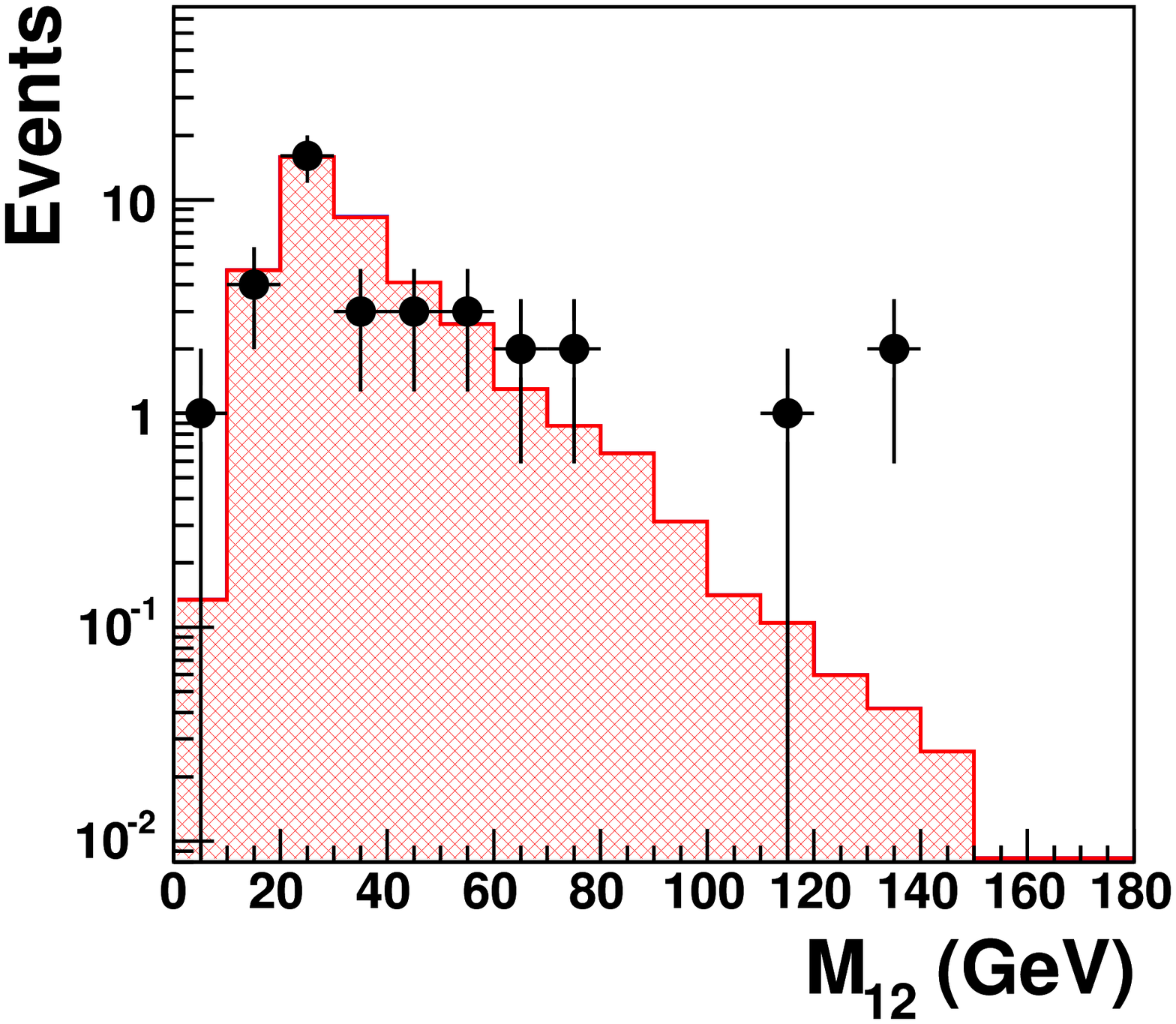}
  \caption{Distributions of the invariant mass of the two highest $P_{T}$ electons in the
	di--electron (left) and tri--electron (right) H1 analyses compared to the
	SM expectation, which is dominated by pair production \cite{multilep}.}
\label{fig:multilep}
\end{figure}

Searches for multi--electron production at high transverse momentum have been previously
carried out by the H1 \cite{multie} and ZEUS experiments \cite{multiezeus}, using the
HERA~I data sample. H1 has additionally studied the production of high $P_{T}$ muon pairs
\cite{multimu}. The main SM process for multi--lepton production in $ep$ collisions is
photon--photon interactions $\gamma\gamma\rightarrow l^{+}l^{-}$, where quasi real
photons radiated from the incoming electron and proton interact to produce a pair of
leptons. High mass events ($M_{1,2} >$~100~GeV) are observed in the in the di--electron
sample of both experiments, and additionally in the H1 tri--electron sample, regions where
the SM expectation is low.

H1 has recently updated the analysis with the new HERA~II data, now exploiting a total
luminosity of 209~pb$^{-1}$ \cite{multilep}. The analysis examines $e\mu$, $\mu\mu$,
$e\mu$, $eee$ and $e\mu\mu$ topologies, searching for events with at least two high
$P_{T}$ electrons or muons. Figure \ref{fig:multilep} shows the invariant mass distribution
of the H1 di--electron (left) and tri--electron (right) samples. The data are
found to be in good overall agreement with the SM, although interesting events are seen
at high masses; in the electron di--electron and tri--electron channels in the HERA~I
data and more recently in the electron--muon channels in the HERA~II data.

The production in $ep$ collisions of a doubly charged Higgs boson $H^{++(--)}$ could be
a source of events containing multiple high $P_{T}$ leptons and the observed high mass
events in the H1 analysis have been investigated in this context \cite{doublehiggs}.
Only one $ee$ event satisfies the additional selection criteria and the
HERA limits on the $H^{++(--)}$ coupling to $ee$ are not competitive to those set by
the OPAL experiment \cite{opalhiggs}. However, limits derived from the HERA data in
the $e\mu$ decay channel of the $H^{++(--)}$ extend to higher masses beyond the reach
of the previous searches performed by the CDF \cite{cdfhiggs} and LEP \cite{lephiggs}
experiments and new constraints were obtained for the $H^{++(--)}$ coupling to $e\tau$.

\subsection{A General Search for New Phenomena}

A model independent general search for deviations from the SM has been performed by H1
using a HERA~I data sample corresponding to an integrated luminosity of 117~pb$^{-1}$
\cite{general}. All high $P_{T}$ final state configurations involving electrons ($e$),
muons ($\mu$), jets ($j$), photons ($\gamma$) or neutrinos ($\nu$) are considered.
All final state configurations containing at least two such objects with $P_{T} >$~20~GeV
in the central region of the detector are investigated and classified into exclusive
event classes, $e$--$j$, $\mu$--$j$--$\nu$, $j$--$j$--$j$ and so on.

\begin{figure}[h]
  \centering
  \includegraphics[height=.6\textheight, angle=90]{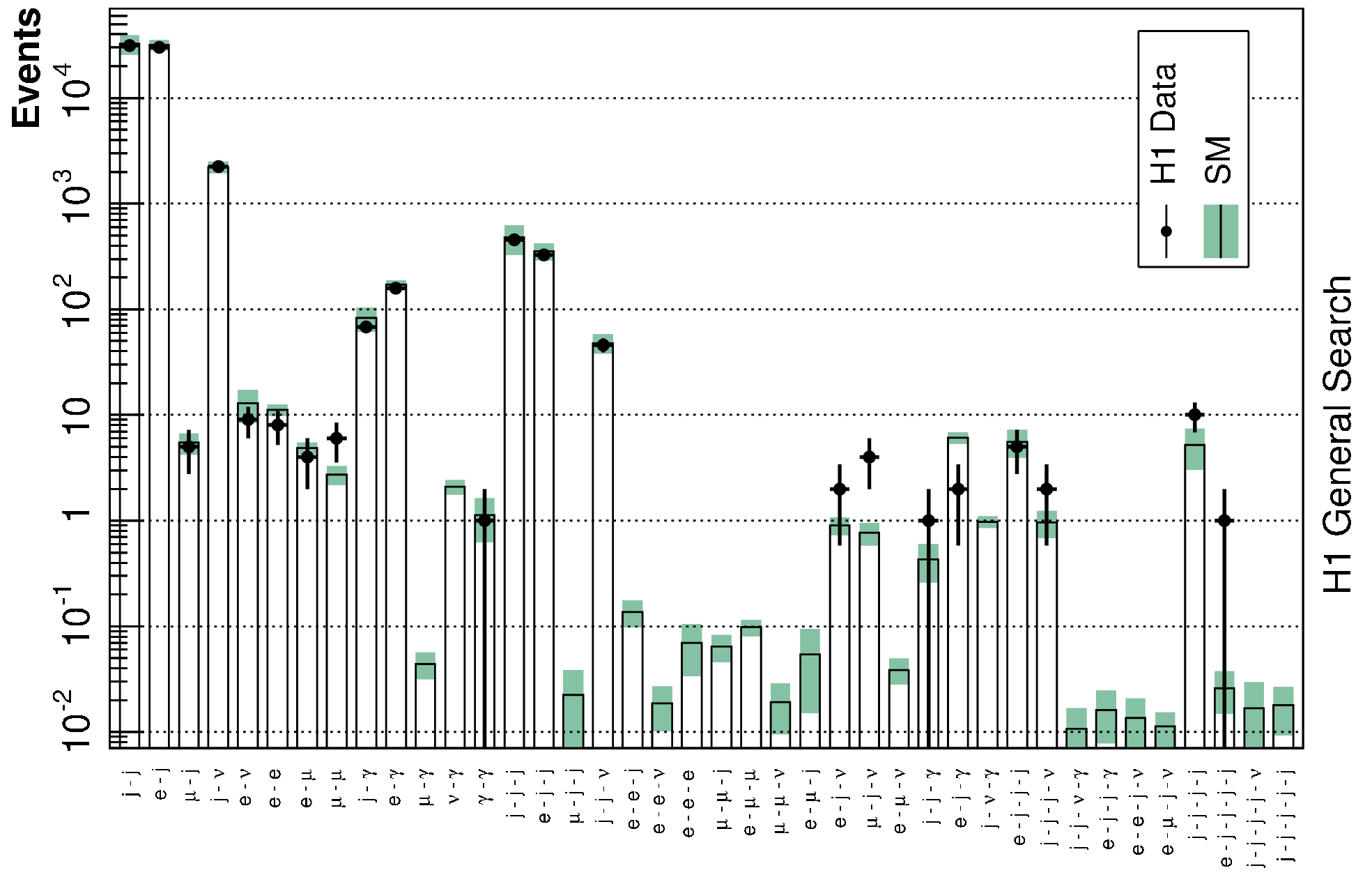}
  \caption{The data and the SM expectation for all event classes with a SM expectation
	greater than 0.01 events in the H1 general search analysis \cite{general}.}
\label{fig:general}
\end{figure}

Data events are found in 22 such event classes and good agreement is observed between
data and the SM expectation in most event classes, as can be seen in figure
\ref{fig:general}. A non--biased statistical method is employed to search for deviations
of the data with respect to the SM. A good agreement is found in all channels,
except in the $\mu$--$j$--$\nu$ event class, where 4 data events are observed compared
to a SM expectation of 0.8~$\pm$~0.2 as previously reported in \cite{isoleph1newwpaper}.
Additionally, in the $e$--$j$--$j$--$j$--$j$ event class 1 event is observed in the data
compared to a SM prediction of 0.026~$\pm$~0.011.

\section{Searches for Physics Beyond the Standard Model}

\subsection{Leptoquark Production and Lepton Flavour Violation}

The $ep$ collsions at HERA provide a unique possibility to investigate
the formation of a new particle coupling to a lepton--quark pair. In the
Buchm\"{u}ller, R\"{u}ckl and Wyler (BRW) classification of such
states, termed ``leptoquarks'', 7 scalar and 7 vector particles are proposed,
all of which can couple to an $eq$ pair and 4 of which can also couple to both
$eq$ and $\nu q$ \cite{brw}.

\begin{figure}[ht]
\begin{minipage}{.5\linewidth}
  \centering
  \includegraphics[width=.97\textwidth]{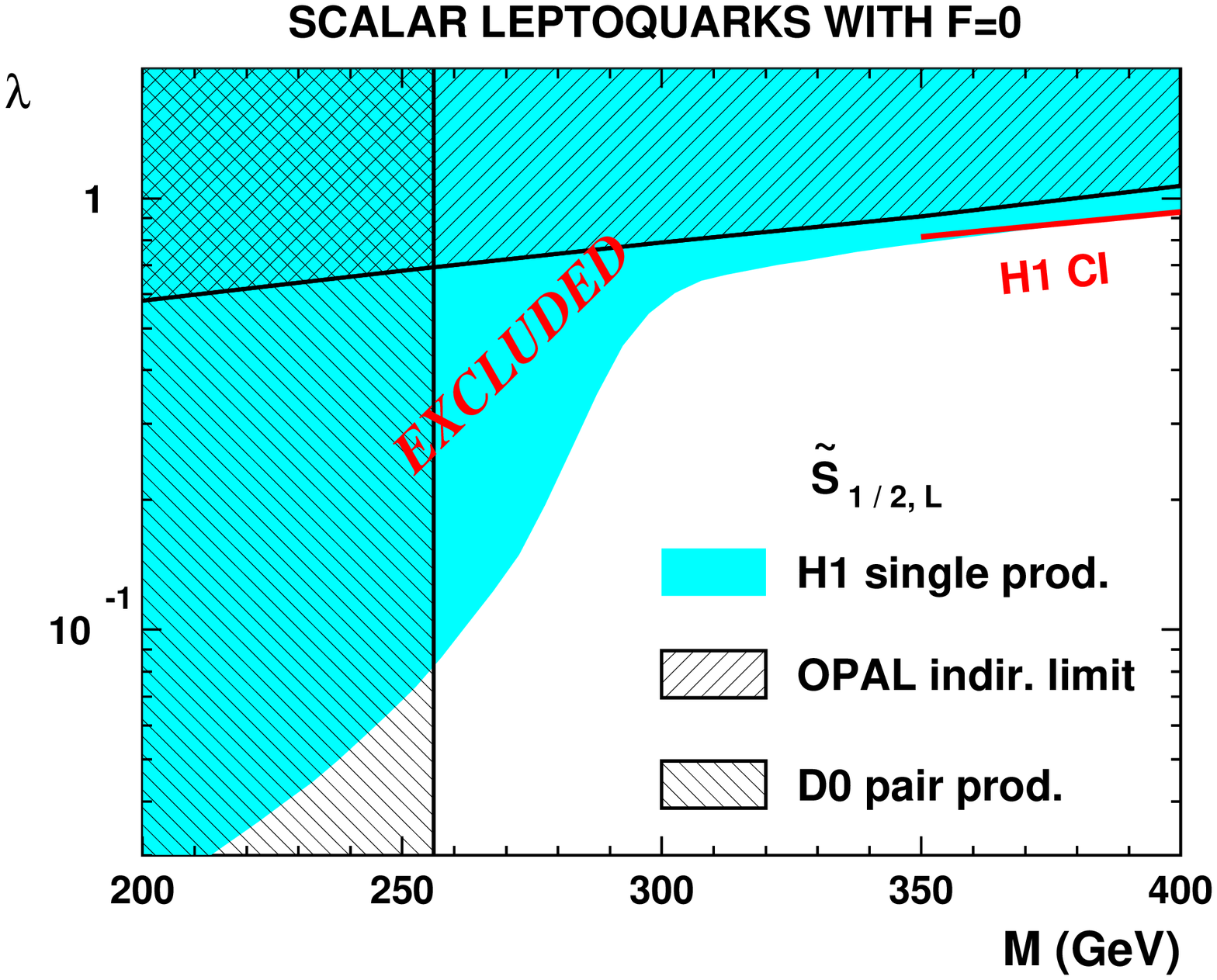}
  \caption{H1 exclusion limits at 95\% CL on the coupling $\lambda$ as a function of
	leptoquark mass for $F=0$ scalar leptoquarks in the framework
	of the BRW model \cite{h1leptoquark}.}
\label{fig:h1lq}
\end{minipage}
\hspace{0.3cm}
\begin{minipage}{.45\linewidth}
  \centering
  \includegraphics[width=.97\textwidth]{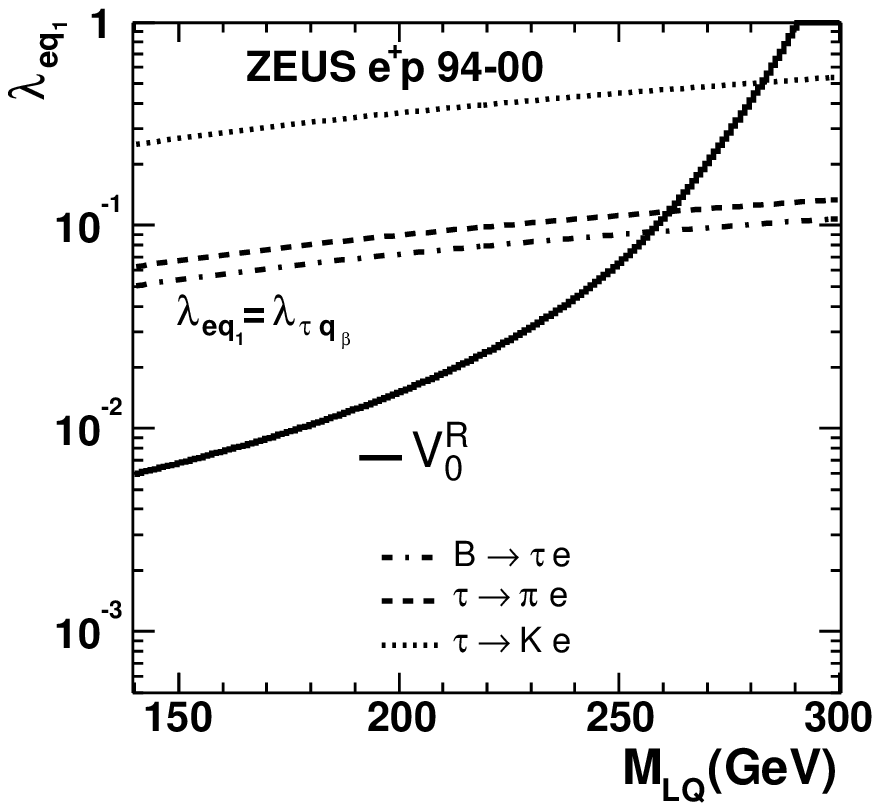}
  \caption{ZEUS exclusion limits at 95\% CL on the coupling $\lambda$ as a function of
	leptoquark mass for the vector leptoquark $V_{0}^{R}$ in the $\tau$ decay channel
	\cite{zeusleptoquarklfv}.}
\label{fig:zeuslfvlq}
\end{minipage}
\end{figure}

A search for first generation leptoquark production, $eq\rightarrow eq$
or $eq\rightarrow \nu q$ has been performed on the HERA~I data by the H1 \cite{h1leptoquark}
and ZEUS \cite{zeusleptoquark} collaborations by looking for deviations in the mass
spectra of neutral current (NC) and charged current (CC) DIS interactions, the major SM
background. No evidence of leptoquark production is observed, and limits on the
14 leptoquark couplings are derived as function of mass, as shown for example from H1
for $F=0$ scalar leptoquarks in figure \ref{fig:h1lq}. In the H1 (ZEUS) analysis,
leptoquark masses up to 325~(386)~GeV are ruled out at 95\% CL for a coupling $\lambda$
of electromagnetic strength ($\lambda$~=~$\sqrt{4 \pi \alpha_{em}} =$~0.3).

A search for lepton flavour violating (LFV) processes mediated by
leptoquark exchange has also been performed by H1 \cite{h1leptoquarklfv} and
ZEUS \cite{zeusleptoquarklfv}, investigating the interactions $eq\rightarrow \mu q$
and $eq \rightarrow \tau q$. No evidence for LFV is observed and, similarly to the
above analyses, limits are derived for the appropriate leptoquark couplings as a function
of mass. An example is shown in figure \ref{fig:zeuslfvlq} from the ZEUS analysis of
the $eq \rightarrow \tau q$ channel using 113~pb$^{-1}$ of $e^{+}p$ data, where the ZEUS
results improve the existing limits from rare $\tau$, $B$ or $K$ decays (see references
given in \cite{zeusleptoquarklfv}).

\subsection{R--Parity Violating SUSY and the Search for Bosonic Stop Decays}

H1 has previously performed a search for squarks ($\tilde{q}$), the scalar
supersymmeteric (SUSY) partners of quarks, in models with R--parity violation
({\mbox{$\not \hspace{-0.1cm} R_{p}$}}) \cite{h1squark}. The search rules out squarks
of all flavours with masses up to 275~GeV for a coupling of electromagnetic strength
in the Minimal Supersymmetric Standard Model (MSSM) framework. A similar analysis
recently performed by the ZEUS collaboration set comparable limits in a search for
stop quark ($\tilde{t}$) production in MSSM {\mbox{$\not \hspace{-0.1cm} R_{p}$}}
SUSY \cite{zeusstop}.

\begin{figure}
  \centering
  \includegraphics[height=.34\textheight]{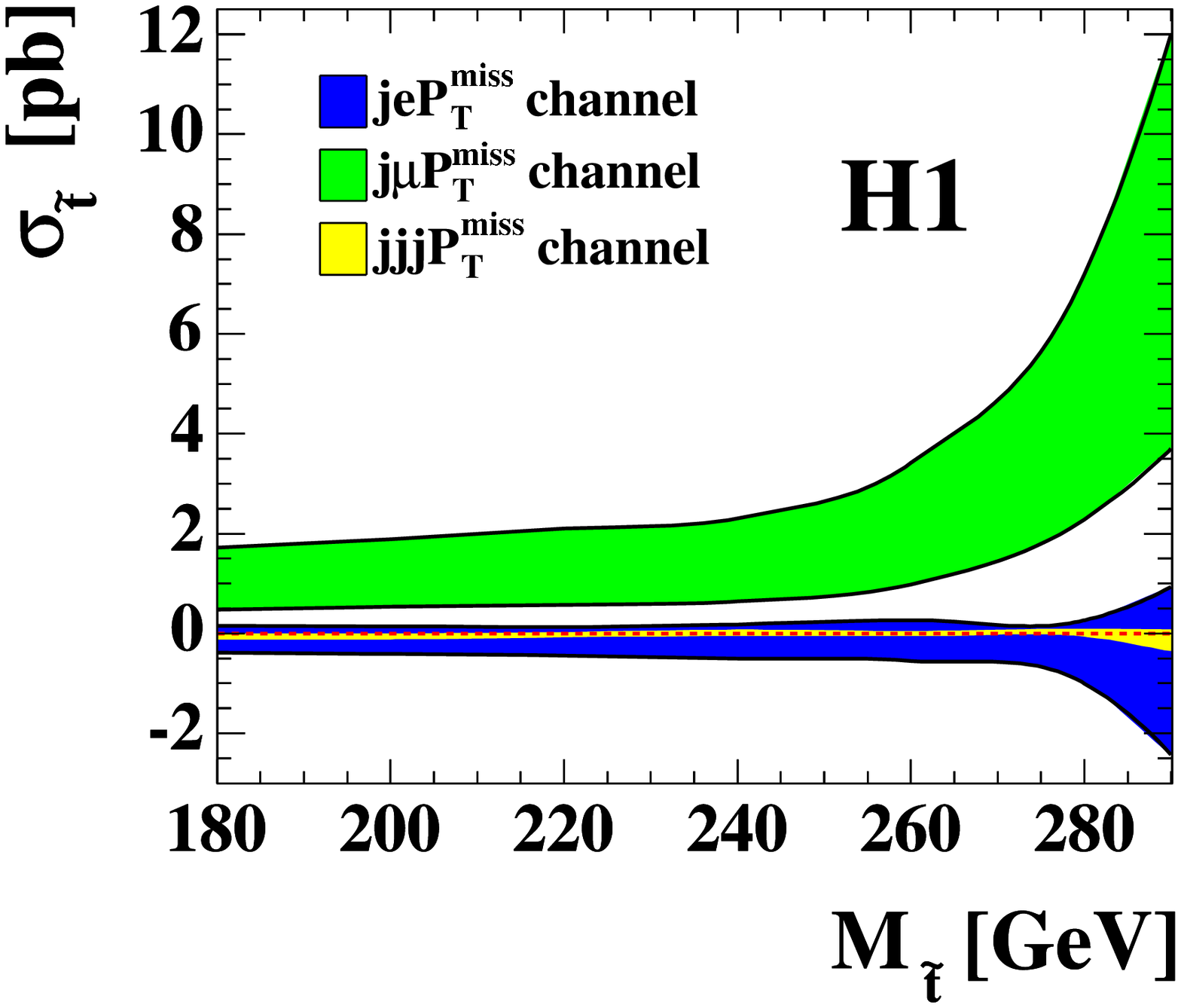}
  \includegraphics[height=.34\textheight]{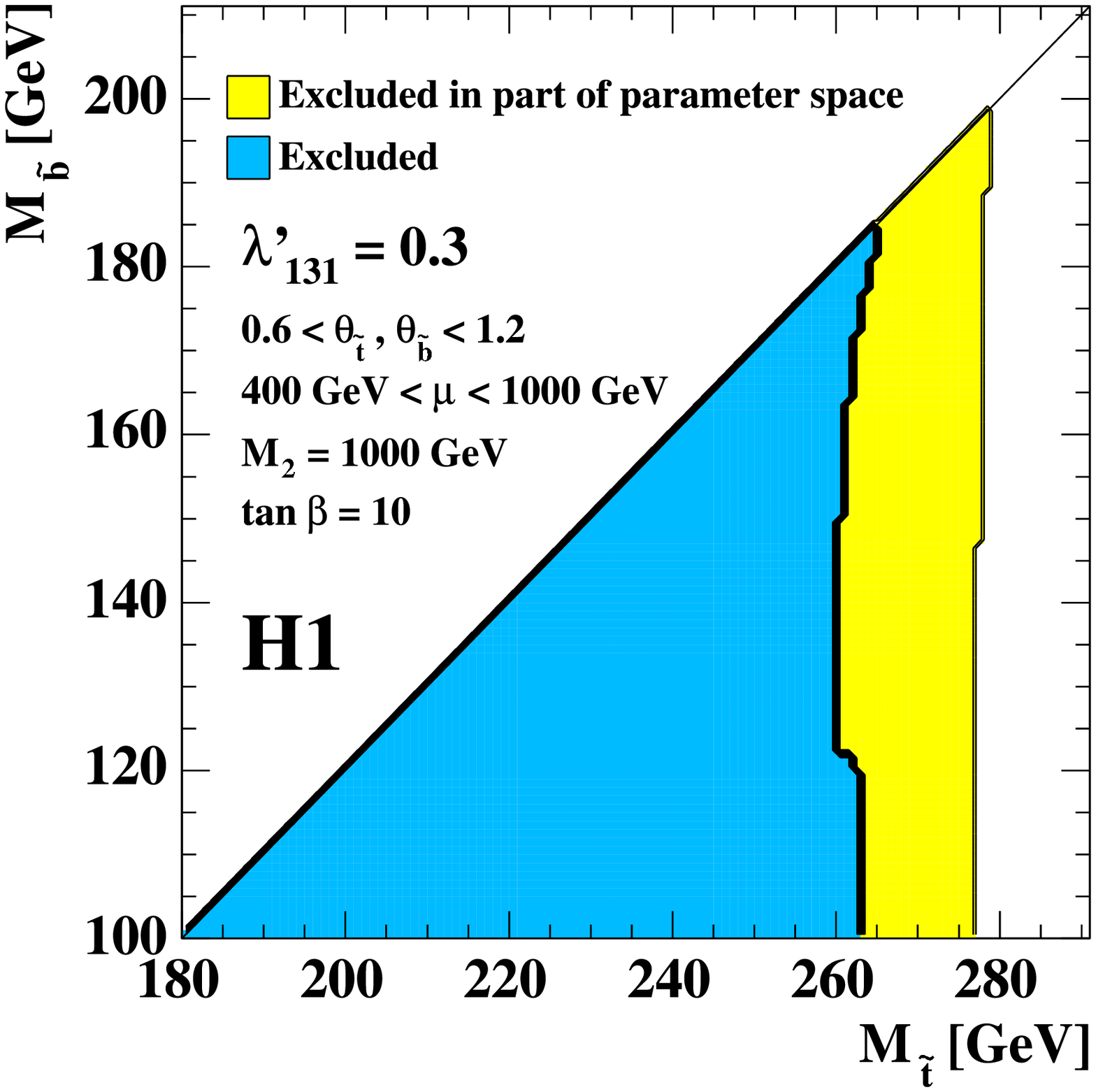}
  \caption{Left: Bands representing the allowed stop cross section regions
	$\sigma_{\tilde{t}} \pm \Delta\sigma_{\tilde{t}}$ as a function of the
	stop mass, obtained from the analysis of each bosonic stop decay channel.
	Right: Exclusion limits at the 95\% CL in the ($M_{\tilde{t}}$,~$M_{\tilde{b}}$)
	plane for $\lambda_{131}^{\prime} =$~0.3 \cite{bosonicstop}.}
\label{fig:bosonicstop}
\end{figure}

In the afforementioned analyses all squarks are assumed to be degenerate.
A complementary analysis, based on the assumption $M_{\tilde{t}} > M_{\tilde{b}}$ has
been performed by the H1 collaboration using $e^{+}p$ data corresponding to an integrated
luminosity of 106~pb$^{-1}$. In this case, the stop produced in
{\mbox{$\not \hspace{-0.1cm} R_{p}$}} fusion via the $\lambda^{\prime}$ coupling decays
bosonically $\tilde{t} \rightarrow \tilde{b}W$ \cite{bosonicstop}.
The {\mbox{$\not \hspace{-0.1cm} R_{p}$}} decay of the sbottom quark
$\tilde{b} \rightarrow d\bar{\nu}_{e}$ and leptonic and hadronic $W$ decays are
considered. The {\mbox{$\not \hspace{-0.1cm} R_{p}$}} decay $\tilde{t} \rightarrow eq$
is also examined for completeness. The bosonic stop decay leads to three different final
state topologies depending on the decay of the $W$ boson; a jet, a lepton (electron or
muon)\footnote{The $W$ decay into $\nu_{\tau}\tau$, where $\tau \rightarrow$ hadrons
$+ \nu$ is not investgated in this analysis.} and missing transverse momentum
($je$\mbox{$\not \hspace{-0.1cm} P_\perp$}~channel~and~$j\mu$\mbox{$\not \hspace{-0.1cm} P_\perp$} channel)
or, if the $W$ decays to jets, three jets and missing transverse momentum
($jjj$\mbox{$\not \hspace{-0.1cm} P_\perp$} channel). A slight excess of events compared
to the SM prediction is observed in the $j\mu$\mbox{$\not \hspace{-0.1cm} P_\perp$} channel,
confirming the results of the dedicated H1 analysis to such a topology \cite{isoleph1newwpaper}.

Assuming the presence of a stop mass $M_{\tilde{t}}$ decaying bosonically, the
observed event yields are used to determine the allowed range for a stop production
cross section $\sigma_{\tilde{t}}$, as illustrated in figure \ref{fig:bosonicstop}
(left). It can be seen that the excess in the
$j\mu$\mbox{$\not \hspace{-0.1cm} P_\perp$} channel is not supported by the other
channels, and hence no evidence for stop production is observed. The results from
the different channels are combined to derive constraints on stop quarks decaying
bosonically in the MSSM. The resulting limits projected on the
($M_{\tilde{t}}$,~$M_{\tilde{b}}$) plane for $\lambda_{131}^{\prime} =$~0.3 are
shown in figure \ref{fig:bosonicstop} (right), where for $M_{\tilde{b}} =$~100~GeV
stop masses up to 275~GeV are ruled out at 95\% CL.

\subsection{Search for Gaugino Production and Events Containing Light Gravitinos}

\begin{figure}[t]
\begin{minipage}{.48\linewidth}
  \centering
  \includegraphics[height=0.27\textheight]{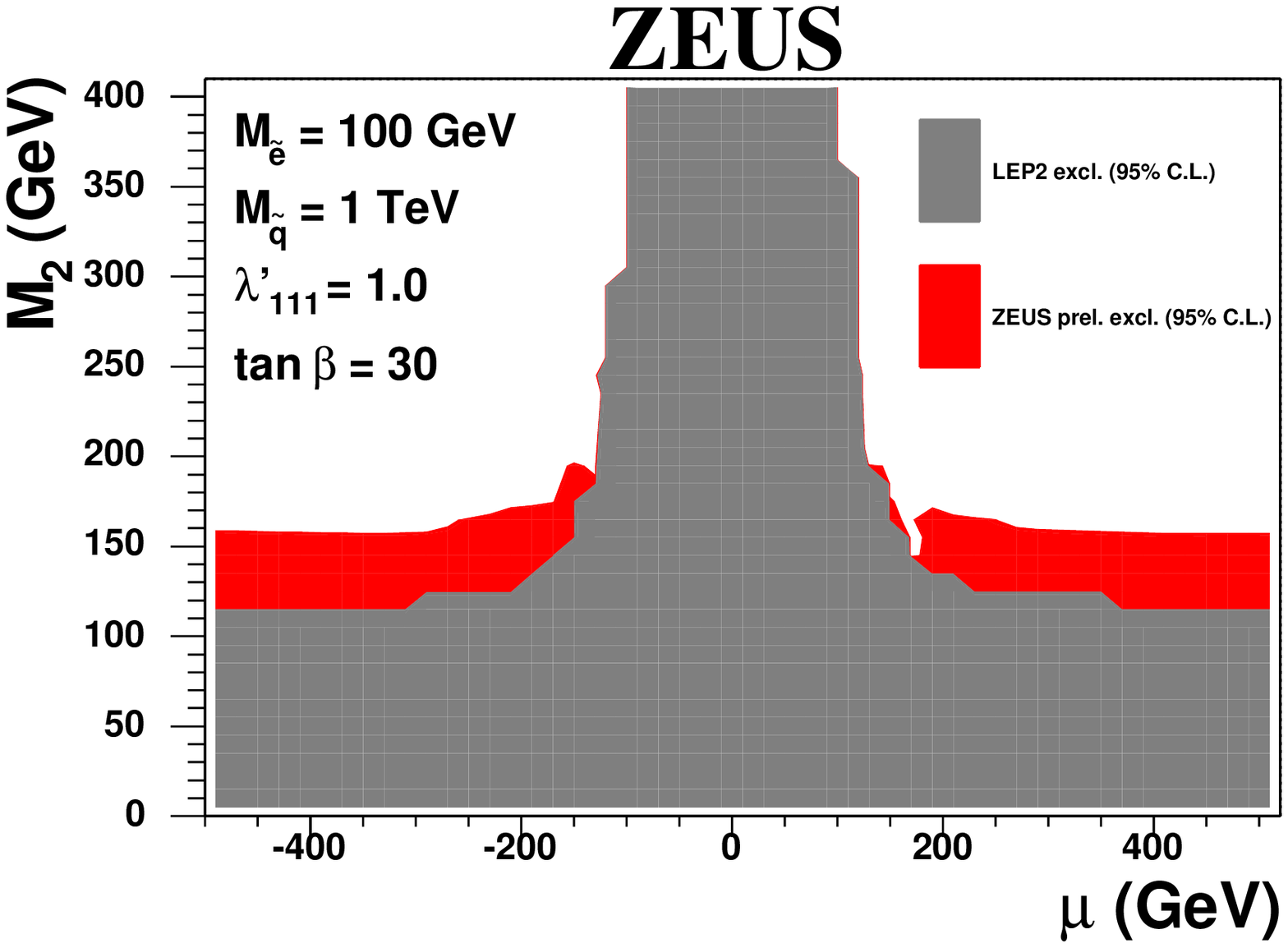}
  \caption{Excluded regions at the 95\% CL in the ($M_{2}$,~$\mu$) plane from the ZEUS search
	for gaugino production \cite{zeusgaugino}, compared to results from the LEP
	experiments ALEPH and DELPHI \cite{lepgaugino}.}
\label{fig:gaugino}
\end{minipage}
\hspace{0.1cm}
\begin{minipage}{.48\linewidth}
  \centering
  \includegraphics[height=0.33\textheight]{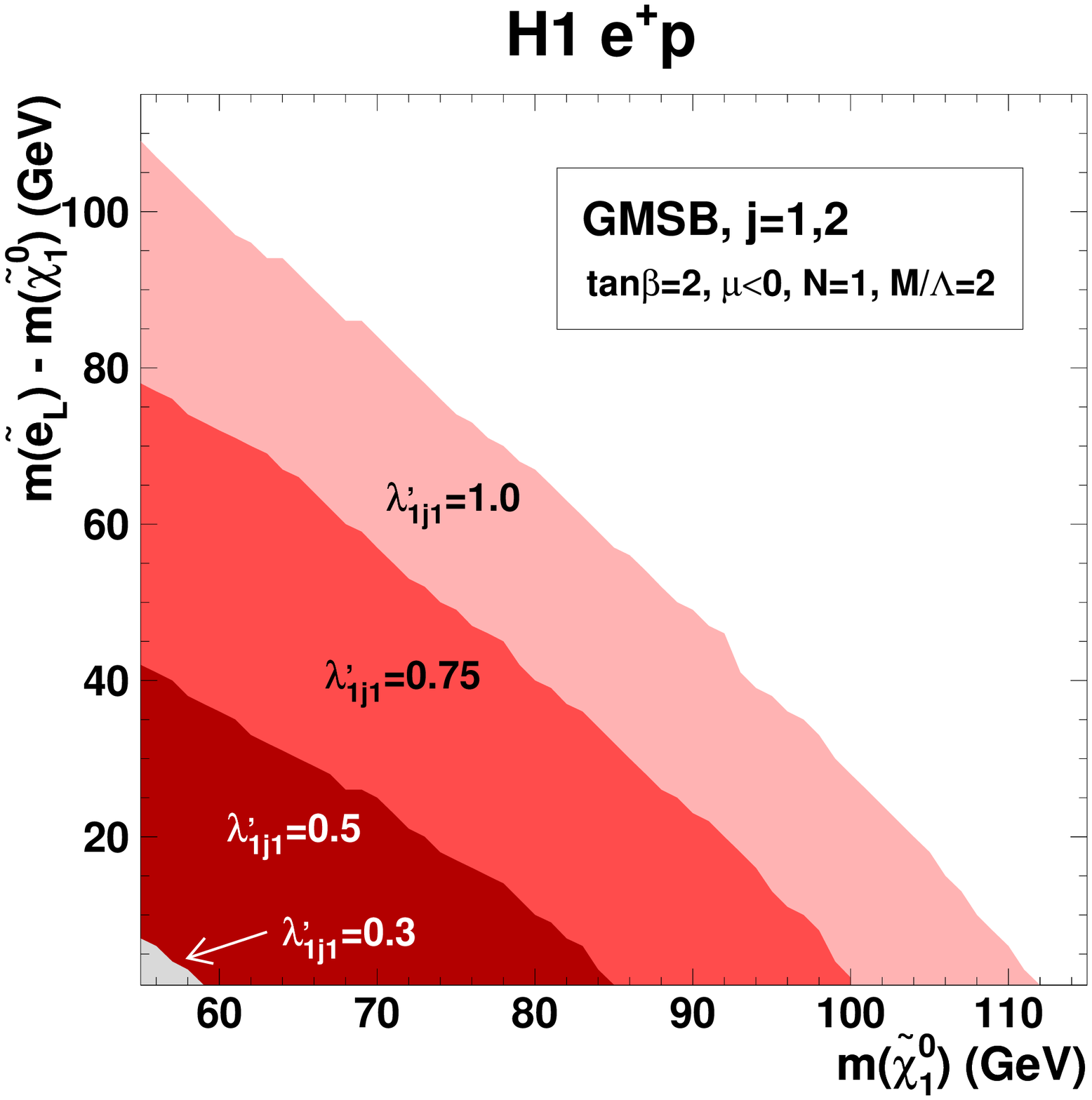}
  \caption{Excluded regions at the 95\% CL in the
	$(m(\tilde{e}_{L}) - m(\tilde{\chi}_{1}^{0})$,~$m(\tilde{\chi}_{1}^{0})$)
	plane from H1 for various values of
	the coupling $\lambda_{1j1}^{\prime}$~($j =$ 1,2) \cite{h1gravitino}.}
\label{fig:gravitino}
\end{minipage}
\end{figure}

If squarks are very heavy, {\mbox{$\not \hspace{-0.1cm} R_{p}$}} SUSY phenomena can manifest
in $ep$ collisions by gaugino production via $t$--channel slepton exchange. A search
for gauginos is performed by the ZEUS collaboration, where the interaction proceeds
via $t$--channel exchange of a selectron and production of a neutralino
$\tilde{\chi}_{1}^{0}$ \cite{zeusgaugino}. The neutralino subsequently cascade decays into
two quarks and an electron, positron or neutrino and hence the main SM background arises
from di--jets in NC and CC DIS. No deviation from the SM is observed, and exclusion
limits in the ($M_{2}$,~$\mu$) plane for gaugino production are derived at 95\% CL,
as shown in figure \ref{fig:gaugino}. These limits extend previous
limits set by the LEP experiments \cite{lepgaugino}.

In other, so called Gauge Mediated Supersymmetry Breaking (GMSB) models, the lightest SUSY
particle (LSP) is the gravitino $\tilde{G}$ and the next--to--lightest SUSY particle (NLSP)
can be the lightest neutralino $\tilde{\chi}_{1}^{0}$, which decays
to the stable gravitino and a photon. This specific decay channel of the neutralino is
investigated by the H1 \cite{h1gravitino} and ZEUS \cite{zeusgravitino}
collaborations as a search for light gravitino production, that is resonant single
neutralino production $\tilde{\chi}_{1}^{0}$
via $t$--channel selectron exchange, $e^{\pm}p \rightarrow \tilde{\chi}_{1}^{0} q^{\prime}$,
where it is assumed the $\tilde{\chi}_{1}^{0}$ is the NLSP and the subsequent
decay $\tilde{\chi}_{1}^{0} \rightarrow \gamma\tilde{G}$ occurs with an
unobservably small lifetime. The experimental signature is thus a photon, a jet from
the struck quark and missing transverse momentum due to the gravitino and hence the
main SM background is due to radiative CC DIS. No significant deviation from the SM is
observed by either H1 or ZEUS and the results are used to derive constraints on GMSB models
for different values of the {\mbox{$\not \hspace{-0.1cm} R_{p}$}} coupling
$\lambda_{1j1}^{\prime}$ for fixed values of the relevant SUSY parameters.
Figure \ref{fig:gravitino} displays excluded regions in the
($m(\tilde{e}_{L}) - m(\tilde{\chi}_{1}^{0})$,~$m(\tilde{\chi}_{1}^{0})$)
plane derived from $e^{+}p$ HERA data from H1 for various values of $\lambda_{1j1}^{\prime}$.
For small mass differences between the neutralino and selectron, neutralino masses up to
112~GeV are ruled out at 95\% CL for an {\mbox{$\not \hspace{-0.1cm} R_{p}$}} coupling
$\lambda_{1j1}^{\prime} =$~1.0. The results presented in this section are the first
constraints from HERA on SUSY models independent of the squark sector.

\section{Conclusions}
Many searches for new physics have been performed at HERA by the H1 and ZEUS
collaborations, some including data from the new HERA II phase. Interesting
events containing isolated leptons and missing $P_{T}$ and multiple high $P_{T}$
leptons at high masses have been observed by H1. No evidence is seen by either
collaboration in searches for a large selection of exotic particles and competitive
limits have been derived on such BSM scenarios.
The continued data taking in the HERA~II phase by both H1 and ZEUS
experiments will hopefully clarify the high $P_{T}$ lepton events seen by H1
and further the search for new physics at HERA.

\end{document}